\documentstyle[12pt,qqa4jart]{article}
\input tcilatex
\QQQ{Language}{
American English
}

\begin{document}

\title{New perturbation theory for nonstationary anharmonic oscillator}
\author{Alexander V. Bogdanov and Ashot S. Gevorkyan}
\date{Institute for High-Performance Computing and Data Bases\\
P/O Box, 71, St-Petersburg, 194291, Russia}
\maketitle

\begin{abstract}
The new perturbation theory for the problem of nonstationary anharmonic
oscillator with polynomial nonstationary perturbation is proposed. As a zero
order approximation the exact wave function of harmonic oscillator with
variable frequency in external field is used. Based on some intrinsic
properties of unperturbed wave function the variational-iterational method
is proposed, that make it possible to correct both the amplitude and the
phase of wave function. As an application the first order correction are
proposed both for wave function and $S$-matrix elements for asymmetric
perturbation potential of type $V(x,\tau )=\alpha (\tau )x^3+\beta (\tau
)x^4.$\ The transition amplitude ''ground state - ground state'' $%
W_{00}\left( \lambda ;\rho \right) $\ is analyzed in detail depending on
perturbation parameter $\lambda $\ (including strong coupling region $%
\lambda $ $\sim 1$)\ and one-dimensional refraction coefficient $\rho $.
\end{abstract}

\section{Introduction}

Every successful effort in investigation of some physical problem in strong
interaction region usually brings new insight in behavior of corresponding
system. Thus, the outstanding results, obtained in the solution of
one-dimensional model problems in quantum field theory, molecular and
solid-state physics $\left[ 1-3\right] $ are of great interest. But the
investigation becomes much more complicated when many-dimensional problems
are approached. In such situation contradictory results are possible even in
calculations of the system eigenenergy $\left[ 4\right] $ .

The situation becomes even more dramatic while dealing with nonstationary
problems with eigenfunctions basis, changing in time. The most important
example of it is given by model of two-dimensional scattering problem, that
is called collinear model for rearrangement $\left[ 5\right] .$ It is one of
the simplest and still realistic descriptions of the three-body reaction of $%
A+(B,C)_n\rightarrow (A,B)_m+C$ type.

This problem can be of principle importance both for the practical
applications in the theory of chemical reaction for moderate energies and
for illuminating the mechanisms of rearrangement in multichannel scattering
theory. It can be also regarded as a good testbed for application of
nonperturbative method in realistic few body problem with all reaction
channels involved.

In our previous paper $\left[ 6\right] $ it was shown, that in the limit $%
\hbar \rightarrow 0$ this problem is effectively reduced to that for
one-dimensional anharmonic oscillators with the frequency $\Omega (\tau )$,
changing in the field of external force $F(\tau )$. Numerous
quantum-chemical calculations of the potential energy surface for most
popular three-particle systems have shown strong anharmonism in normal
coordinate even for principle quantum number as low, as $n\geq 3\div 4$.

Here we propose the method, that seems to be a natural development of the
ideas of stationary perturbation theory of $\left[ 3\right] $ for the case
of nonstationary coefficients of a zero-order equation. The method is of
variational/iteration type and is essentially backed by main ideas of
singular perturbation theory - one can expand only such parts of the
solution, that do not bring singularities, and if one can not find pattern
for singular part of the solution it is possible to expand it in the
exponential.

Our final goal would be derivation of the scattering matrix representation,
that makes it possible the simultaneous computation of the probabilities for
all pertinent scattering mechanisms, taking into account close coupling of
many channels, resonance effects and chaotic behavior in intermediate state.

The perturbation theory proposed might be especially interesting for
calculation of wave packet of scattering system, since harmonic
approximation, that exactly propagates Gaussian wave packets, is a natural
starting point of that theory $\left[ 7\right] $.

The suggested development covers also two of the principle questions of the
quantum theory. First one is connected to the possibility of study of the
quantum system behavior in the region of close coupling. For that purpose it
is necessary to build up a perturbation theory, that do not need a knowledge
of the spectrum of unperturbed system. Also there are several such
approaches (see i.e. [3]), they are not generalized to nonstationary case
and so are not suited for scattering problems. Our method exactly serves for
that purpose.

Second question is closely connected to the limits of quasiclassical
description of the quantum system evolution. The traditional WKB approach,
as it is well known, is limited for scattering applications. At the same
time there is famous path integral approach that makes it possible to
describe exactly quantum dynamics in terms of classical trajectories. So the
basic question is how to improve the WKB approximation to describe all the
effects, for which traditional approach fails. Our investigations makes it
possible to develope the representation of the scattering theory via the
classical trajectories and so build up the intermediate approximate
solutions between exact and primitive WKB forms.

It is well known, that nonstationary form of the equation for quantum
oscillator by a special transformations can be reduced to stationary form
[10], that can be solved exactly. In our case unfortunately the use of such
approach is not very effective, sine after such transformation the
perurbation becomes so complicated, that analytic computation of the
perturbation matrix elements would be unrealistic.

\section{Formulation of the problem}

Let us consider the Schroedinger equation for one-dimensional nonstationary
anharmonic oscillator

\begin{equation}
\stackrel{\wedge }{L}\Psi +\lambda V\left( x,\tau \right) \Psi =0,\qquad
-\infty <x,\tau <+\infty , 
\end{equation}

\begin{equation}
\stackrel{\wedge }{L}=i\partial _\tau +\frac 12\partial _x^2-(\frac 12\Omega
^2(\tau )x^2-F(\tau )x), 
\end{equation}

with $\Psi $ being the oscillator wave function, $\lambda $ being the
dimensionless coupling parameter, and the perturbation potential $V(x,\tau )$%
, oscillator frequency $\Omega (\tau )$ and the external force $F(\tau )$
having the following asymptotic behavior

\begin{equation}
V_{\pm }(x)=\stackunder{\tau \rightarrow \pm \infty }{\lim }V(x,\tau
),\qquad \Omega _{in(out)}=\stackunder{\tau \rightarrow \pm \infty }{\lim }%
\Omega (\tau ),\qquad \stackunder{\tau \rightarrow \pm \infty }{\lim }F(\tau
)=0. 
\end{equation}

Let us assume, that solution of $\left( 2.1\right) $ satisfies the following
boundary and initial conditions

\begin{equation}
\begin{array}{c}
\stackunder{x\rightarrow \pm \infty }{\lim }\Psi (x,\tau )=0,\qquad 
\stackunder{x\rightarrow \pm \infty }{\lim }\Psi (x,\tau )\partial _x\Psi
(x,\tau )=0, \\  \\ 
\stackunder{\tau \rightarrow -\infty }{\lim }\Psi (x,\tau )=\Psi
_{in}(n;\Omega _{in},x)\exp \left[ -i\left( n+\frac 12\right) \Omega
_{in}\tau \right] , 
\end{array}
\end{equation}

were $\Psi _{in}(n;\Omega _{in},x)$ is a function of stationary anharmonic
oscillator of initial channel.

The physical meaning of those conditions is obvious - the decay of
probability and its flow at infinity - and it corresponds to the formulation
of the physical problems, discussed in Introduction. When perturbation is
absent, i.e. with $V=0$, the equation

\begin{equation}
\stackrel{\wedge }{L}\Psi _0(n;x,\tau )=0, 
\end{equation}

with boundary conditions $\left( 2.3\right) $, $\left( 2.4\right) $ has an
exact solution (see i.e. $\left[ 8\right] $ )

\begin{equation}
\Psi _0(n;x,\tau )=f_0(n;x,\tau )=K(n;\tau )\exp \left( a_1(\tau )y+a_2(\tau
)y^2\right) H_n(y), 
\end{equation}

$$
K(n;\tau )=\left( \frac{(\Omega _{in}/\pi )^{1/2}}{2^nn!|\zeta (\tau )|}%
\right) ^{1/2}\exp \left( i\int\limits_{-\infty }^\tau L_{cl}(\tau )d\tau
-i\int\limits_{-\infty }^\tau E(n;\tau )d\tau \right) , 
$$

$$
a_1(\tau )=i\frac{\dot \eta |\zeta |}{(\Omega _{in})^{1/2}},\qquad a_2(\tau
)=\frac 12\left( i\frac{|\zeta ||\dot \zeta |}{\Omega _{in}}-1\right)
,\qquad \dot \eta =d\eta /d\tau 
$$

\begin{equation}
y=\left( \Omega _{in}\right) ^{1/2}\frac{x-\eta (\tau )}{|\zeta (\tau )|}%
,\qquad |\dot \zeta |=\frac{d|\zeta |}{d\tau } 
\end{equation}
$$
L_{cl}=\frac 12\left( \dot \eta \right) ^2-\frac 12\Omega ^2\eta ^2+F\eta
,\qquad E(n;\tau )=\frac{\Omega _{in}}{|\zeta (\tau )|^2}(n+1/2), 
$$

with constant $\Omega _{in}$ being the initial frequency, $n$ being the
principle quantum number, $E(n;\tau )$ being the adiabatically changing $n$%
-the energy level and the functions $\zeta \left( \tau \right) $, $\eta
\left( \tau \right) $ satisfy the equations

\begin{equation}
\ddot \zeta +\Omega ^2(\tau )\zeta =0,\qquad \ddot \eta +\Omega ^2(\tau
)\eta =F(\tau ), 
\end{equation}

with following asymptotic and initial conditions

$$
\zeta _{+}\left( \tau \right) =\lim \limits_{\tau \rightarrow \infty }\zeta
\left( \tau \right) =c_1\exp \left( i\Omega _{out}\tau \right) -c_2\exp
\left( -i\Omega _{out}\tau \right) , 
$$

\begin{equation}
\zeta _{-}(\tau )=\stackunder{\tau \rightarrow -\infty }{\lim }\zeta (\tau
)=\exp \left( i\Omega _{in}\tau \right) ,\qquad |c_1|^2-|c_2|^2=1, 
\end{equation}

$$
\eta \left( -\infty \right) =\dot \eta \left( -\infty \right) =0. 
$$

It is well known, that solution of the second equation $\left( 2.8\right) $
can be constructed on the base of solution of corresponding homogeneous
equation

\begin{equation}
\eta \left( \tau \right) =\frac 1{\sqrt{2\Omega _{in}}}\left[ \zeta \left(
\tau \right) d^{*}\left( \tau \right) +\zeta ^{*}\left( \tau \right) d\left(
\tau \right) \right] ,\qquad d\left( \tau \right) =\frac i{\sqrt{\Omega _{in}%
}}\int\limits_{-\infty }^\tau d\tau ^{^{\prime }}\zeta \left( \tau
^{^{\prime }}\right) F\left( \tau ^{^{\prime }}\right) . 
\end{equation}

Semiclassical type analysis gives the following form of solution of $\left(
2.1\right) $

\begin{equation}
\Psi ^{(+)}(n;x,\tau )=f(n;x,\tau )\exp \left( -\Phi (n;x,\tau )\right) , 
\end{equation}

with $\Psi ^{(+)}$ being the total wavefunction, that is being developed
from the $n$-the asymptotic excited state at $\tau \rightarrow -\infty $.

Substitution of $\left( 2.11\right) $ into $\left( 2.1\right) $ gives the
following equation for unknown functions $\Phi $ and $\,f$

\begin{equation}
\left( i\partial _\tau \Phi -\frac 12\left( \partial _x\Phi \right) ^2+\frac
12\partial _x^2\Phi -\lambda V(x,\tau )\right) f-\left( \stackrel{\wedge }{L}%
f-(\partial _x\Phi )(\partial _xf)\right) =0. 
\end{equation}

For further investigation of equation $\left( 2.12\right) $ it is convenient
to expand $f(n;x,\tau )$ and $\Phi (n;x,\tau )$ functions into power series
over $\lambda $

\begin{equation}
\Phi (n;x,\tau )=\sum\limits_{k=0}^\infty \lambda ^k\Phi _k(n;x,\tau
),\qquad \Phi _0(n;x,\tau )=0, 
\end{equation}

\begin{equation}
f(n;x,\tau )=\sum\limits_{k=0}^\infty \lambda ^kf_k(n;x,\tau ). 
\end{equation}

As in all exponential approximation approaches one term of expansion of $%
\Phi (n;x,\tau )$ takes into account infinite number of terms of standard
perturbation theory. But the more we try to take into account in the
exponential, the more difficult would be the equations for individual
members. In our case there is an optimal choice, that makes it possible to
formulate additional condition for $\left( 2.13\right) ,$ $\left(
2.14\right) $, and so to determine both functions $\Phi $ and $f$ in (2.11)
without ambiguity. Let us start from the equation for ''$k$''-the correction

\begin{equation}
i\partial _\tau \Phi _k+\frac 12\partial _x^2\Phi _k+q_k^{(1)}(x,\tau
)-f_0^{-1}\left( \stackrel{\wedge }{L}f_k-(\partial _x\Phi _k)(\partial
_xf_0)+q_k^{(2)}(x,\tau )\right) =0, 
\end{equation}

with

$$
q_k^{(1)}=-\frac 12\sum\limits_{m=1}^{k-1}(\partial _x\Phi _m)(\partial
_x\Phi _{k-m}),\qquad k\geq 2, 
$$

\begin{equation}
q_k^{(2)}(x,\tau )=-\sum\limits_{m=1}^{k-1}[f_m(i\partial _\tau \Phi
_{k-m}+\frac 12\partial _x^2\Phi _{k-m}-\delta _{k-m,1}V- 
\end{equation}

$$
-\frac 12\sum\limits_{l=1}^{k-m-1}(\partial _x\Phi _l)(\partial _x\Phi
_{k-m-l}))-(\partial _x\Phi _m)(\partial _xf_{k-m})],\qquad k\geq 2. 
$$

In case of $k=1$ one has simple relations

\begin{equation}
q_1^{(1)}(x,\tau )=-V(x,\tau ),\qquad q_1^{(2)}(x,\tau )=0. 
\end{equation}

Note, that the eigenenergies are not preserved in our problem and so the
corrections to them in general give no useful information about the system.
Thus the problem is reduced to the solution of equation $\left( 2.15\right) $
i.e. to determination of correction to wavefunction of anharmonic oscillator.

\section{Construction of the solution}

Before solving the equation $\left( 2.15\right) $ let us note two important
points

a) The equation is not correctly formulated, as there are two unknown
functions in one equation,

b) There is, in general, at least one singular member in this equation. It
can be explained most easily from analysis of $\left( 2.16\right) $, taking
into account, that $f_0(n;x,\tau )$ includes the Hermitian polynomial. This
difficulty is readily solved if the perturbation is of polynomial form, as
we would suppose in future.

Such form of perturbation does not seriously influence our discussion, as
any smooth perturbation can be approximated by a finite part of its Taylor
expansion.

As in every exponential perturbation theory one can solve both above
problems by choosing additional condition for equation $\left( 2.15\right) $%
. Let us start from equations of the first order in perturbation. It is
convenient to choose

\begin{equation}
i\partial _\tau \Phi _1+\frac 12\partial _x^2\Phi _1+q_1^{(1)}(n;x,\tau
)+Q_1(n;x,\tau )=0, 
\end{equation}

\begin{equation}
\stackrel{\wedge }{L}f_1=(\partial _x\Phi _1)(\partial _xf_0)-Q_1(n;x,\tau
)f_{0,} 
\end{equation}

with $Q_1(n;x,\tau )$ being yet unknown function. With additional condition
of $\Phi _1(n;x,\tau )$ being nonsingular correction, one has that $\Phi
_1(n;x,\tau )$, $Q_1(n;x,\tau )$ and all the terms in $\left( 3.2\right) $
are polynomials of the same order, as $q_1^{(1)}(n;x,\tau )$. By expanding
the functions of eq. $\left( 3.2\right) $ in a series of Hermitian
polynomials and supposing the absence of the members of an order, higher
than $n$, one can correctly determine the coefficients of polynomial $%
Q_1(n;x,\tau ).$ So, it is clear that both above difficulties are thus
overcome and the system of equations $\left( 3.1\right) ,\left( 3.2\right) $
is quite correct. The procedure is readily generalized to any order. As for $%
k\geq 2\;q_k^{(2)}(n;x,\tau )\neq 0$, so one has

\begin{equation}
Q_k(n;x,\tau )=\{(\partial _x\Phi _k)(\partial _xf_0)-\stackrel{\wedge }{L}%
f_k-\sum\limits_{m=1}^{k-1}(f_m[(i\partial _\tau \Phi _{k-m})+\frac
12\partial _x^2\Phi _{k-m}- 
\end{equation}

$$
-\delta _{k-m,1}V-\frac 12\sum\limits_{l=1}^{k-m-1}(\partial _x\Phi
_l)(\partial _x\Phi _{k-m-l})]-(\partial _xf_{k-m})(\partial _x\Phi
_m))\}f_0^{-1}. 
$$

Taking into account, that expression in square brackets is equal to $Q_{k-l}$%
, eq. $\left( 3.3\right) $ is written in the following form

\begin{equation}
Q_k(n;x,\tau )=\left\{ (\partial _x\Phi _k)(\partial _xf_0)-\stackrel{\wedge 
}{L}f_k-\sum\limits_{m=1}^{k-1}\left( f_mQ_{k-m}-(\partial _xf_m)(\partial
_x\Phi _{k-m})\right) \right\} f_0^{-1}, 
\end{equation}

and for determination of $f_k$ one has the equation

\begin{equation}
\stackrel{\wedge }{L}f_k=\sum\limits_{m=1}^{k-1}\left( (\partial
_xf_m)(\partial _x\Phi _{k-m})-f_mQ_{k-m}\right) . 
\end{equation}

And again if one supposes, that $f_m$, $\Phi _m$ and $Q_m$ for $m\leq k$ are
polynomials, then $Q_k$ and $f_k$ are easily determined and it is possible
to find the equation for $\Phi _k$

\begin{equation}
i\partial _\tau \Phi _k+\frac 12\partial _x^2\Phi _k+q_k^{(1)}(n;x,\tau
)+Q_k(n;x,\tau )=0. 
\end{equation}

Note, that for polynomial $Q_k(n;x,\tau )$ the solution of $\left(
3.6\right) $ is also polynomial and only high order terms of perturbation
potential $V(x,\tau )$ are important for determination of $\Phi _m$ and $f_m$%
.

\section{Calculation of wavefunction to the first order of perturbation
theory}

Let us study in more detail the case of asymmetric polynomial perturbation

\begin{equation}
V(x,\tau )=\alpha (\tau )x^3+\beta (\tau )x^4=\sum\limits_{m=0}^4b_m(\tau
)y^m, 
\end{equation}

with coefficients $b_m(\tau )$ of the form

$$
b_0(\tau )=\eta ^3(\beta \eta +\alpha ),\;b_1(\tau )=\Omega _{in}^{-1/2}\eta
^2|\zeta |(4\beta \eta +3\alpha ), 
$$

$$
b_2(\tau )=3\Omega _{in}^{-1}\eta |\zeta |^2(2\beta \eta +\alpha
),\;b_3(\tau )=\Omega _{in}^{-3/2}|\zeta |^3(4\beta \eta +\alpha ), 
$$

\begin{equation}
b_4(\tau )=\Omega _{in}^{-2}\beta |\zeta |^4, 
\end{equation}

and assumption, that $\alpha (\tau )$ and $\beta (\tau )$ are slowly varying
functions of $\tau $.

Before passing to the solution of $\left( 3.1\right) $ , $\left( 3.2\right) $
in accordance to previous discussion let us rewrite $\Phi _1$ and $Q_1$ in
the following form

\begin{equation}
\Phi _1(n;x,\tau )=\sum\limits_{k=0}^4v_k(n;\tau )y^k,\;Q_1(n;x,\tau
)=\sum\limits_{k=0}^4\sigma _k(n;\tau )y^k. 
\end{equation}

Thus, the additional condition, introduced for regularization of
perturbation approach, can be expressed in the form

$$
f_0(n;x,\tau )\sum\limits_{k=0}^4\sigma _k(\tau )y^k=\kappa _0((2a_2(\tau
)y+a_1(\tau ))f_0(n;x,\tau )+2nK(n;\tau )\times 
$$

\begin{equation}
\times K^{-1}(n-1;\tau )f_0(n-1;x,\tau ))\sum\limits_{k=1}^4kv_k(\tau
)y^{k-1}-\stackrel{\wedge }{L}f_1,\qquad \kappa _0=\frac{\Omega _{in}}{%
|\zeta (\tau )|^2}. 
\end{equation}

Writing down $f_0(n;x,\tau )$ in explicit form and equating on both sides of 
$\left( 4.4\right) $ coefficients of the polynomials of the same order we
may determine the coefficients in polynomial $Q_1(n;x,\tau ):$

$$
\sigma _0(n;\tau )=\kappa _0(a_1v_1+2nv_2+2n(n-1)v_4),\quad \sigma _1(n;\tau
)=2\kappa _0(a_2v_1+a_1v_2+\frac 32nv_3), 
$$

$$
\sigma _2(n;\tau )=4\kappa _0(a_2v+\frac 34a_1v_3+nv_4),\quad \sigma
_3(n;\tau )=6\kappa _0(a_2v_3+\frac 23a_1v_4), 
$$

\begin{equation}
\sigma _4(n;\tau )=8\kappa _0a_2v_4. 
\end{equation}

Substituting $\left( 4.5\right) $ in $\left( 3.1\right) $ one gets a system
of nonuniform linear differential equations for determination of
coefficients in the correction $\Phi _1(n;x,\tau ):$

\begin{equation}
i\dot v_j-c_j(\tau )v_j-d_j(\tau )=0,\quad \dot v_j=dv_j(\tau )/d\tau
,\qquad j=0,1,2,3,4. 
\end{equation}

with functions $c_j(\tau )$ and $d_j(\tau )$ given by

$$
c_4(\tau )=4\kappa _0,\quad \;c_3(\tau )=3\kappa _0,\quad c_2(\tau )=2\kappa
_0,\quad \ c_1(\tau )=\kappa _0,\quad \;c_0(\tau )=0, 
$$

\begin{equation}
d_4(\tau )=b_4,\quad \;d_3(\tau )=b_3,\qquad d_2(\tau )=b_2+2(2n+3)\kappa
_0v_4, 
\end{equation}

$$
d_1(\tau )=b_1+3(n+1)\kappa _0v_3,\qquad d_0(\tau )=b_0+(2n+1)\kappa
_0v_2+2n(n-1)\kappa _0v_4. 
$$

Initial conditions for the system $\left( 4.5\right) $ are

\begin{equation}
v_0^{-}=0,\;\dot v_0^{-}=-\,id_0,\;v_j^{-}=-\,\frac{d_j^{-}}{c_j^{-}},\qquad
\;j=1,2,3,4. 
\end{equation}

Note, that ''$-$'' parameters correspond to the limit $\tau \rightarrow
-\infty $ . It is clear that solution of $\left( 4.6\right) $ must start
from $j=4$ . In such a way, solution of each equation is presented in the
following form

\begin{equation}
v_j(\tau )=G_j\left( \tau ,-\infty \right) \left[
v_j^{-}-i\int\limits_{-\infty }^\tau G_j\left( -\infty ,\tau ^{\prime
}\right) d_j\left( \tau ^{\prime }\right) d\tau ^{\prime }\right],\qquad
\;j=0,...,4, 
\end{equation}

with $G_j\left( \tau ,\tau ^{\prime }\right) $ being the evolution operator
of pertinent homogeneous equation

\begin{equation}
G_j\left( \tau ,\tau ^{\prime }\right) =G_j^{-1}\left( \tau ^{\prime },\tau
\right) =\exp \left[ -\,ij\int\limits_{\tau ^{\prime }}^\tau \kappa _0\left(
\tau ^{\prime \prime }\right) d\tau ^{\prime \prime }\right] ,\; 
\end{equation}

Taking into account conditions $\left( 4.5\right) $ equation $\left(
4.4\right) $ can be represented in the following form 
\begin{equation}
\stackrel{\wedge }{L}f_1\left( n;x,\tau \right) -\sum\limits_{j=1}^4%
\stackrel{\_}{e}_j\left( n;\tau \right) f_0\left( n-j;x,\tau \right) =0, 
\end{equation}

with

$$
\stackrel{\_}{e}_4\left( n;\tau \right) =2\kappa _0v_4\left[ \frac{n!}{%
\left( n-4\right) !}\right] ^{1/2}G_4\left( -\infty ,\tau \right) , 
$$

$$
\stackrel{\_}{e}_3\left( n;\tau \right) =3\kappa _0v_3\left[ \frac{n!}{%
2\left( n-3\right) !}\right] ^{1/2}G_3\left( -\infty ,\tau \right) , 
$$

$$
\stackrel{\_}{e}_2=\kappa _0\left[ 2\left( 2n-3\right) v_4+v_2\right] \left[ 
\frac{n!}{\left( n-2\right) !}\right] ^{1/2}G_2\left( -\infty ,\tau \right)
, 
$$

\begin{equation}
\stackrel{\_}{e}_1\left( n;\tau \right) =\kappa _0\left[ 3\left( n-1\right)
v_3+2v_1\right] \left[ \frac{n!}{2\left( n-1\right) !}\right]
^{1/2}G_1\left( -\infty ,\tau \right) . 
\end{equation}

In all cases, when $n-j\prec 0$, $\stackrel{\_}{e}_j$ are equal to zero. The
solution of $\left( 4.11\right) $ is naturally represented in such a way

\begin{equation}
f_1\left( n;x,\tau \right) =\sum\limits_{j=1}^4\stackrel{\_}{w}_j\left(
n;\tau \right) f_0\left( n-j;x,\tau \right) ,\;\stackrel{\_}{w}_j\left(
n;\tau \right) =G_j\left( -\infty ,\tau \right) w_j\left( n;\tau \right) . 
\end{equation}

Substituting $\left( 4.13\right) $ into $\left( 4.11\right) $ one gets the
linear first-order equations for coefficients $w_j\left( \tau \right) $

\begin{equation}
i\dot w_j-j\kappa _0\left( \tau \right) w_j-e_j\left( n;\tau \right)
=0,\;e_j\left( n;\tau \right) =\stackrel{\_}{e}_j\left( n;\tau \right)
G_j\left( \tau ,-\infty \right) , 
\end{equation}

with initial conditions

\begin{equation}
w_j^{-}=\lim _{\tau \rightarrow -\infty }w_j\left( n;\tau \right) =-\frac{%
e_j\left( n,-\infty \right) }{j\kappa _0\left( -\infty \right) },\qquad
\;j=1,...,4. 
\end{equation}

Solution of this equation is obvious,

\begin{equation}
w_j\left( \tau \right) =G_j\left( \tau ,-\infty \right) \left[
w_j^{-}-i\int\limits_{-\infty }^\tau d\tau ^{\prime }G_j\left( -\infty ,\tau
^{\prime }\right) e_j\left( \tau ^{\prime }\right) \right] . 
\end{equation}

Solutions for higher orders of expansion are constructed in the same way.

\section{Calculation of the transition $S$-matrix for nonstationary
anharmonic oscillator}

The fact, that unharmonic oscillator wave functions, determined above, are
nonstationary, opens new possibilities in view of our previous result [6].
It was shown there, that scattering in three-body collinear system is
effectively reduced to the evolution of anharmonic oscillator in external
field. And so the above wave functions can be used for computation of the
scattering matrix. It is not evident how to calculate the scattering matrix
elements via the oscillator functions. It is possible to show, using the
development of [5], that $S-$matrix elements for scattering problems can be
represented by one-dimensional integral in the form, similar to standard
representation [9] of nonstationary $S-$matrix:

\begin{equation}
S_{mn}=\lim _{\tau \rightarrow +\infty }\left\langle \psi _f^{*}\left(
m;x,\tau \right) \psi ^{+}\left( n;x,\tau \right) \right\rangle ,\qquad
\left\langle ....\right\rangle =\int\limits_{-\infty }^\infty ...dx, 
\end{equation}

with $\psi _f\left( m;x,\tau \right) $ being the asymptotic wave function of
final state. Let us have a look at the approximations of exact and
asymptotic wave functions. For nonstationary wave function $\psi ^{\left(
+\right) }\left( n;x,\tau \right) $ with the help of $\left( 2.11\right) $
to the first order of perturbation theory over $\lambda $ one has

\begin{equation}
\psi ^{\left( +\right) }\left( n;x,\tau \right) =\left[ f_\lambda \left(
n;x,\tau \right) +\lambda f_\lambda ^1\left( n;x,\tau \right) \right] \exp
\left[ -\lambda \sum\limits_{l=1}^4v_l\left( \tau \right) y^l\right]
+O\left( \lambda ^2\right) , 
\end{equation}

$$
f_\lambda \left( n;x,\tau \right) =\exp \left[ -\lambda v_0\left( \tau
\right) \right] f_0\left( n;x,\tau \right) ,\;f_\lambda ^1\left( n;x,\tau
\right) =\exp \left[ -\lambda v_0\left( \tau \right) \right] f_1\left(
n;x,\tau \right) . 
$$

Note, that functions $f_0\left( n;x,\tau \right) $ and $f_1\left( n;x,\tau
\right) $ are determined by the formulae $\left( 2.6\right) -\left(
2.7\right) $ and $\left( 4.13\right) $. Expanding the exponential function
in $\left( 5.2\right) $ one has

\begin{equation}
\psi ^{\left( +\right) }\left( n;x,\tau \right) =f_\lambda \left( n;x,\tau
\right) +\lambda \left[ f_\lambda ^1\left( n;x,\tau \right) -f_\lambda
\left( n;x,\tau \right) \sum\limits_{l=1}^4v_l\left( \tau \right) y^l\right]
+O\left( \lambda ^2\right) . 
\end{equation}

Now one can use well known formula for Hermitian polynomials $xH_m\left(
x\right) =1/2H_{m+1}\left( x\right) +mH_{m-1}\left( x\right) $ and expand in 
$\left( 5.3\right) $ the expressions of type $y^jf_\lambda \left( n;x,\tau
\right) $ into the series over the Hermitian polynomials. So one gets

\begin{equation}
\begin{array}{c}
\psi ^{\left( +\right) }\left( n;x,\tau \right) =f_\lambda \left( n;x,\tau
\right) + \\  
\\ 
+\lambda \left[ \sum\limits_{j=1}^4\stackrel{\_}{w}_l\left( n;\tau \right)
f_\lambda \left( n-l;x,\tau \right) -\sum\limits_{p=-4}^4\stackrel{\_}{u}%
_p\left( n;\tau \right) f_\lambda \left( n-p;x,\tau \right) \right] +O\left(
\lambda ^2\right) , 
\end{array}
\end{equation}

with

\begin{equation}
\stackrel{\_}{u}_p\left( n;\tau \right) =G_p\left( \tau ,-\infty \right)
u_p\left( n;\tau \right) ,\quad u_p\left( n;\tau \right) =\left[ \frac{%
\left( n-p\right) !}{2^pn!}\right] ^{1/2}\chi _p\left( n;\tau \right) . 
\end{equation}

As to $\chi _p\left( n;\tau \right) $ functions, they are given by the
following expressions

$$
\chi _{-4}\left( n;\tau \right) =\frac 1{2^4}v_4\left( \tau \right) ,\qquad
\chi _{-3}\left( n;\tau \right) =\frac 1{2^3}v_3\left( \tau \right) , 
$$

$$
\chi _{-2}\left( n;\tau \right) =\frac 12\left( n+\frac 32\right) v_4\left(
\tau \right) +\frac 1{2^2}v_2\left( \tau \right) ,\qquad \chi _{-1}\left(
n;\tau \right) =\frac 34\left( n+1\right) v_3\left( \tau \right) +\frac
12v_1\left( \tau \right) , 
$$

\begin{equation}
\chi _0\left( n;\tau \right) =\frac 32\left( n^2+n+\frac 12\right) v_4\left(
\tau \right) +\left( n+\frac 12\right) v_2\left( \tau \right) , 
\end{equation}

$$
\chi _4\left( n;\tau \right) =n\left( n-1\right) \left( n-2\right) \left(
n-3\right) v_4\left( \tau \right) ,\qquad \chi _3\left( n;\tau \right)
=n\left( n-1\right) \left( n-2\right) v_3\left( \tau \right) 
$$

$$
\chi _2\left( n;\tau \right) =3n\left( n-1\right) ^2v_4\left( \tau \right)
+n\left( n-1\right) v_2\left( \tau \right) ,\qquad \chi _1\left( n;\tau
\right) =\frac 32n^2v_3\left( \tau \right) +nv_1\left( \tau \right) . 
$$

In the same way we can get the representations for asymptotic states of
anharmonic oscillator. So for $(out)$ state one has

\begin{equation}
\begin{array}{c}
\psi _f\left( m;x,\tau \right) =\varphi _f^0\left( m;x,\tau \right) + \\  
\\ 
+\lambda \left[ \sum\limits_{l=1}^4\stackrel{\_}{w}_l^f\left( m;\tau \right)
\varphi _f^0\left( m-l;x,\tau \right) -\!\sum\limits_{p=-4}^4\stackrel{\_}{u}%
_p^f\left( m;\tau \right) \varphi _f^0\left( m-p;x,\tau \right) \right]
+O\left( \lambda ^2\right) . 
\end{array}
\end{equation}

Function $\varphi _f^0\left( m;x,\tau \right) $ in the above expression is
an $(out)$ asymptotic state of harmonic oscillator

\begin{equation}
\varphi _f^0\left( m;x,\tau \right) =\varphi _f^0\left( m;\Omega
_{out};x\right) \exp \left[ -i\left( m+\frac 12\right) \Omega _{out}\tau
\right] , 
\end{equation}

\begin{equation}
\varphi _f^0\left( m;\Omega ;x\right) =\left[ \frac{\left( \Omega /\pi
\right) ^{1/2}}{2^mm!}\right] ^{1/2}\exp \left( -\frac 12\Omega x^2\right)
H_m\left( \sqrt{\Omega }x\right) . 
\end{equation}

In the above expressions $\stackrel{\_}{w}_l^f\left( m;\tau \right) $ and $%
\stackrel{\_}{u}_p^f\left( m;\tau \right) $ functions are given by

\begin{equation}
\stackrel{\_}{w}_l^f\left( m;\tau \right) =w_l^f\left( m\right) \exp \left(
-il\Omega _{out}\tau \right) , 
\end{equation}

\begin{equation}
\stackrel{\_}{u}_p^f\left( n;\tau \right) =u_p^f\left( n\right) \exp \left(
-ip\Omega _{out}\tau \right) . 
\end{equation}

As to the coefficients $w_l^f\left( m\right) $ and $u_p^f\left( n\right) $,
they can be obtained at once from $\left( 4.16\right) $ and $\left(
5.7\right) -\left( 5.8\right) $ in the limit $\tau \rightarrow -\infty $
after substitution $w_j^{-}\rightarrow w_j^f$ and $v_j^{-}\rightarrow v_j^f$%
. Now taking into account $\left( 5.5\right) $ and $\left( 5.10\right) $ and
using $\left( 5.3\right) $ and $\left( 5.6\right) $ in $\left( 5.1\right) $
one gets

\begin{equation}
S_{mn}\left( \lambda \right) =\left( S_{mn}^0+\lambda \left[
S_{mn}^1+S_{mn}^2\right] \right) \exp \left[ -\lambda v_0^{\left( +\right)
}\right] +O\left( \lambda ^2\right) , 
\end{equation}

with

\begin{equation}
S_{mn}^1=\sum\limits_{l=1}^4w_l^f\left( m\right) S_{\left( m-l\right)
n}^0-\sum\limits_{p=-4}^4u_p^f\left( m\right) S_{\left( m-p\right) n}^0, 
\end{equation}

\begin{equation}
S_{mn}^2=\sum\limits_{l=1}^4w_l^{\left( +\right) }\left( m\right) S_{m\left(
n-l\right) }^0-\sum\limits_{p=-4}^4u_p^{\left( +\right) }\left( n\right)
S_{m\left( n-p\right) }^0, 
\end{equation}

\begin{equation}
S_{mn}^0=\lim \limits_{\tau \rightarrow +\infty }\left\langle \left( \varphi
_f^0\left( m;x,\tau \right) \right) ^{*}f_0\left( n;x,\tau \right)
\right\rangle . 
\end{equation}

Now, starting from $\left( 5.12\right) -\left( 5.15\right) ,$ it is not
difficult to get the analytic expressions for transition probabilities for
anharmonic oscillator,

\begin{equation}
W_{mn}\left( \lambda \right) =\exp \left[ -2\lambda v_0^{\left( +\right)
}\right] \left[ 1+2\lambda Re\left( \frac{S_{mn}^1+S_{mn}^2}{S_{mn}^0}%
\right) \right] W_{mn}^0+O\left( \lambda ^2\right) , 
\end{equation}

\begin{equation}
W_{mn}^0=|S_{mn}^0|^2=\lim \limits_{\lambda \rightarrow 0}W_{mn}\left(
\lambda \right) . 
\end{equation}

In the expressions above, the value $S_{mn}^0$ is a matrix element of $S$%
-matrix for transitions in harmonic oscillator with variable frequency $%
\Omega \left( \tau \right) $ in the external field. The matrix element in $%
\left( 5.15\right) $ is calculated via the generating function $\left[
6\right] $ (see also $\left[ 10\right] $). So we can give here only the
final result

\begin{equation}
W_{mn}^0=\left( \frac{1-\rho }{m!n!}\right) ^{1/2}|H_{mn}\left(
y_1,y_2\right) |^2\exp \left[ -\nu \left( 1-\sqrt{\rho }\cos 2\theta \right)
\right] . 
\end{equation}

Here $H_{mn}\left( y_1,y_2\right) $ is the Hermitian polynomial of two
variables $\left[ 10\right] $ with

\begin{equation}
y_1=\sqrt{\nu \left( 1-\rho \right) }e^{i\theta },\qquad y_2=-\sqrt{\nu }%
\left( e^{-i\theta }-\sqrt{\rho }e^{i\theta }\right) ,\qquad \theta =\frac
12\left( \delta _1+\delta _2\right) -\beta . 
\end{equation}

$\nu ,\rho ,\delta _j$ and $\beta $ parameters are determined from solution
of the classical problem for harmonic oscillator $\left( 2.3\right) ,\left(
2.9\right) $ and are given by following expressions:

\begin{equation}
c_1=e^{i\delta _1}\sqrt{\frac{\Omega _{in}}{\Omega _{out}}}\frac 1{\left(
1-\rho \right) ^{1/2}},\qquad c_2=e^{i\delta _2}\sqrt{\frac{\Omega _{in}}{%
\Omega _{out}}}\left( \frac \rho {1-\rho }\right) ^{1/2}, 
\end{equation}

$$
\rho =\left| \frac{c_2}{c_1}\right| ^2,\qquad d=\lim \limits_{\tau
\rightarrow +\infty }d\left( \tau \right) =\sqrt{\nu }e^{i\beta }. 
$$

Note, that in the expression proposed both for $S$-matrix (5.1) and for
transition probability $W_{mn}$ (5.16) we used only the first members of the
expansion of exponentials in Taylor series over coordinate $y$. Such
approximations for wave function and transition probability are effectively
used for many interesting applications. In some cases, when anharmonic
perturbation substantially change spectrum of the problem and thus our
expansion bases, the leading terms of perturbation must be taken into
account.

It is clear, that for scattering problems unharmonic perturbation is very
important for higher excited states, especially for rearrangement procceses.
At the same time in some situations for rearrangement processes even for
ground state the effect of perturbation can be substantial. To show it let
us discuss one particular case.

\section{Application to ''ground state-ground state'' transition}

We shall demonstrate an application of the method proposed to most simple
situation - the perturbation of parametric harmonic oscillator by symmetric
potential $\beta (\tau )\,x^4$, with $\beta \left( \tau \right) $ being
adiabatically changing functions with boundary values $\beta \left( -\infty
\right) =0$ and $\beta \left( \infty \right) =\beta ^{+}\neq 0$. Note, that
transition probability for unperturbed oscillator can be obtained from $%
\left( 5.16\right) $ with $\nu =0$, or by Taylor expansion of generating
function of corresponding $S$-matrix $\left[ 10\right] $

\begin{equation}
W_{mn}^{\left( 0\right) }=|S_{mn}^{\left( 0\right) }|^2=\frac{n_{<}!}{n_{>}!}%
\sqrt{1-\rho }|P_{\left( n_{<}+n_{_{>}}\right) /2}^{\left(
n_{>}-n_{<}\right) /2}\left( \sqrt{1-\rho }\right) |^2, 
\end{equation}

\begin{equation}
S_{mn}^{\left( 0\right) }\left( \rho \right) =\frac 1{\sqrt{m!n!}}\left\{
\partial _{z_1}^m\partial _{z_2}^nI\left( z_1,z_2;\rho \right) \right\}
_{z_1=z_2=0}, 
\end{equation}

were generating function $I$ is equal

\begin{equation}
I\left( z_1,z_2;\rho \right) =\left( 1-\rho \right) ^{1/4}\exp \left\{ \frac
12\left[ \sqrt{\rho }\left( z_1^2-z_2^2\right) +2\sqrt{1-\rho }z_1z_2\right]
\right\} , 
\end{equation}

with $n_{<}=\min \left( m,n\right) ,\;n_{>}=\max \left( m,n\right) $, and $%
P_n^m\left( x\right) $ being the associated Legendre polynomial.

For many reasons one of most important parameters of the problem is $%
W_{00}\left( \lambda ;\rho \right) $, that measures the probability of
change of initial ground state to final ground state. So we shall discuss it
in some details. From $\left( 6.1\right) -\left( 6.3\right) $ one gets

\begin{equation}
W_{00}\left( \lambda ;\rho \right) =\exp \left[ -\lambda v_0^{\left(
+\right) }\right] \left[ 1-2Re\Lambda \left( \rho \right) \right]
W_{00}^{\left( 0\right) }\left( \rho \right) , 
\end{equation}

\begin{equation}
\Lambda \left( \rho \right) =\left[ S_{00}^{\left( 0\right) }\right]
^{-1}\sum\limits_{k=0}^2\left( u_{-2k}^fS_{2k,0}^{\left( 0\right)
}+u_{-2k}^{\left( +\right) }S_{0,2k}^{\left( 0\right) }\right) , 
\end{equation}

with matrix elements $S_{2k,0}^{\left( 0\right) }$ and $S_{02,k}^{\left(
0\right) }$ given by $\left( 6.2\right) -\left( 6.3\right) $,

\begin{equation}
S_{00}^{\left( 0\right) }=\left( 1-\rho \right) ^{1/4},\qquad S_{20}^{\left(
0\right) }=-S_{02}^{\left( 0\right) }=\frac 1{\sqrt{2!}}\sqrt{\rho }%
S_{00}^{\left( 0\right) },\qquad \;S_{40}^{\left( 0\right) }=S_{04}^{\left(
0\right) }=\frac 1{\sqrt{4!}}\rho S_{00}^{\left( 0\right) }. 
\end{equation}

For determination of $u_{-2k}^f$ and $u_{-2k}^{\left( +\right) }$
coefficients, that are important for $\Lambda \left( \rho \right) $
dependence and so the dependence of scattering amplitude, one must find
coefficients $v_j^f$ and $v_j^{\left( +\right) }=v_j\left( \infty \right) $.
After integration of $\left( 4.9\right) $ by parts and taking into account
adiabatic dependence $\beta \left( \tau \right) $ one gets

\begin{equation}
v_j\simeq -\frac{d_j}{jk_0\left( \tau \right) }+\frac 1jG_j\left( \tau
;-\infty \right) \int\limits_{-\infty }^\tau d\tau ^{\prime }G_j\left(
-\infty ,\tau ^{\prime }\right) \frac{\dot d_j\left( \tau ^{\prime }\right) 
}{k_0\left( \tau ^{\prime }\right) },\qquad j=1,2,3,4. 
\end{equation}

Now the coefficients $v_j^{\left( +\right) }$ are determined from $\left(
6.7\right) $ taking into account $\left( 4.7\right) -\left( 4.8\right) $ via
integration by parts and averaging over fast oscillations

\begin{equation}
v_4^{\left( +\right) }\simeq -\frac{\beta ^{+}}{4\stackrel{\_}{k}_0^3}%
,\qquad v_2^{\left( +\right) }\simeq \frac{3\beta ^{+}}{4\stackrel{\_}{k}_0^3%
},\qquad v_0^{\left( +\right) }\simeq \frac{3\beta ^{+}}{8\stackrel{\_}{k}%
_0^3}, 
\end{equation}

$$
\stackrel{\_}{k}_0=\Omega _{in}/\left( |c_1|^2+|c_2|^2\right) . 
$$

In the same way, for coefficients $v_j^f$ one has

\begin{equation}
v_4^f=-\frac{d_4^f}{c_4^f}=-\frac 14\frac{\beta ^{+}}{\Omega _{out}^3}%
,\qquad v_2^f=-\frac{d_2^f}{c_2^f}=\frac 34\frac{\beta ^{+}}{\Omega _{out}^3}%
,\qquad v_0^f=0, 
\end{equation}
$$
c_j^f=j\Omega _{out.} 
$$

Now using $\left( 5.5\right) -\left( 5.7\right) $ from $\left( 6.8\right)
-\left( 6.9\right) $ one gets the expressions for coefficients $u_{-2k}^{+}$
and $u_{-2k}^f$ in the form

\begin{equation}
u_0^f=\frac 3{16}\left( \frac{\beta ^{+}}{\Omega _{out}^3}\right) ,\qquad
u_0^{+}=\frac 3{16}\frac{\beta ^{+}}{\Omega _{out}^3}\left( \frac{1+\rho }{%
1-\rho }\right) ^3, 
\end{equation}

$$
u_{-2}^f=u_{-2}^{+}=0,\qquad u_{-4}^f=-\frac 14\sqrt{\frac 32}\frac{\beta
^{+}}{\Omega _{out}^3},\qquad u_{-4}^{+}=-\frac 14\sqrt{\frac 32}\frac{\beta
^{+}}{\Omega _{out}^3}\left( \frac{1+\rho }{1-\rho }\right) ^3. 
$$

The formula $\left( 6.4\right) $ with the results above, gives the final
expression for $W_{00}$

\begin{equation}
W_{00}\left( \lambda ;\rho \right) \simeq \sqrt{1-\rho }\left\{ 1-\lambda
\left[ 1-v_0^{+}\left( \rho \right) \right] \left( 1-\frac 13\rho \right)
\right\} \exp \left[ -\lambda v_0^{+}\left( \rho \right) \right] , 
\end{equation}
with new notations

$$
v_0^{+}\left( \rho \right) =\frac{\Omega _{out}^3}{\beta ^{+}}%
v_0^{(+)}=\left( \frac{1+\rho }{1-\rho }\right) ^3,\qquad \lambda
\rightarrow \frac{3\beta ^{+}}{8\Omega _{out}^3}\lambda . 
$$

As it was shown in $\left[ 10\right] ,$ parameter $\rho $, that measures the
excitation of classical oscillator, corresponds to quantum mechanical
refection coefficient of the particle with momentum $k\left( x\right)
=\Omega \left( x\right) $. That makes it possible to use well known results
from quantum mechanics for $\rho $.

As can be seen from fig.1 anharmonic oscillator unlike the harmonic
parametric one in the limit $\rho \rightarrow 0$ have a transition
probability non equal to one. More than that as it is seen from (6.11) the
dependence of transition probability over $\lambda $ is regular, that makes
it possible to use the proposed formula up to the values of $\lambda $ of
the order of unity.

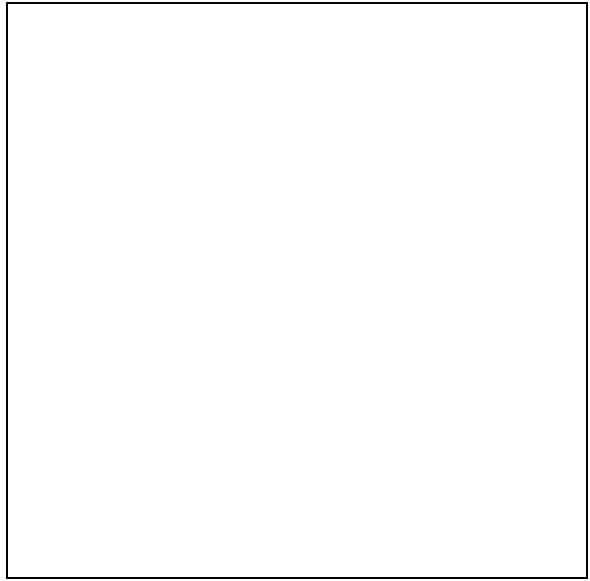
\FRAME{dtbpFUX}{3.0294in}{3.0095in}{0pt}{\Qcb{Fig. 1. The dependence of
''ground state-ground state'' transition probability over reflection
coefficient $\rho $ and compliny constant $\lambda $.}}{}{random.gif}{%
\special{language "Scientific Word";type "GRAPHIC";maintain-aspect-ratio
TRUE;display "USEDEF";valid_file "F";width 3.0294in;height 3.0095in;depth
0pt;cropleft "0";croptop "0.9986";cropright "1";cropbottom "0";filename
'RANDOM.GIF';file-properties "XNPEU";}}

\section{Conclusions}

Many important problems in theoretical and mathematical physics are reduced
to the solution of the equation of nonstationary unharmonic oscillator with
different sets of $\left( in\right) $ and $\left( out\right) $ states. The
use of standard approaches to such problem meets two basic difficulties:

a) Opposite to the stationary situation we do not have the fixed basic set
for perturbed wave function;

b) In nonstationary situation the dimensionless perturbation parameter
changes with time and can become not small in strong coupling region.

In this communication, as a generalization of nonlinearization method [3],
we propose the way to overcome the above difficulties. The perturbation
theory is constructed on exact wave functions for quantum harmonic
oscillator with variable frequency in external field as basis state. This
makes it possible, due to some unique intrinsic properties of those
solutions, to work out the system of two linear equations, that determine
the first corrections both to the amplitude and to the phase of total wave
function. It should be noted also, that $n$-th order correction is also
determined by two independent differential equations, that are obtained
after $n$ iterations.

Very important computational property of our approach is connected to the
fact, that any order correction is constructed on the bases of finite number
of wave vectors. That makes it possible the analytical calculation of the
corrections even to the transition operator.

The possibility of simultaneous correction of the phase and amplitude of the
wave functions, as in a case of stationary problem, gives a regular method
of investigation of strong coupling region, where perturbation is strong and
small parameter is absent.

The approach is well suited for use in numerous application, of which we
point out only some most important:

* corrections to path integral representation of the propagator in field and
scattering theories,

* account of kinetics in some problems of solid state and condensed matter
physics,

* anharmonic corrections for scattering matrix in collinear model of
rearrangement collisions,

* anharmonic corrections for the propagator in the wave packet dynamics
approach to molecular scattering.

We shall discuss those problems in detail in our future publications.

Note, that analysis of proposed expressions show, that account for
antisymmetric terms in potential expansion can cause the nonadiabatic
behavior and so nonanalytic dependence of $W_{00}\left( \lambda ;\rho
\right) $ over $\lambda $. Depending on the form of perturbation ''ground
state - ground state'' transition probability might have in such a case
several maxima.

Authors gratefully acknowledge the helpful discussions with Dr. Yu.
Gorbachev.

\end{document}